\documentclass{vgtc}                          

\ifpdf
  \pdfoutput=1\relax                   
  \usepackage{graphicx}                
  \DeclareGraphicsExtensions{.pdf,.png,.jpg,.jpeg} 
\else
  \usepackage{graphicx}                
  \DeclareGraphicsExtensions{.eps}     
\fi%

\graphicspath{{figures/}{pictures/}{images/}{./}} 

\usepackage{microtype}                 
\PassOptionsToPackage{warn}{textcomp}  
\usepackage{textcomp}                  
\usepackage{mathptmx}                  
\usepackage{times}                     
\usepackage{cite}                      
\usepackage{tabu}                      
\usepackage{booktabs}                  

\onlineid{0}

\vgtccategory{Research}

\vgtcinsertpkg

\usepackage{enumitem}
\usepackage{subcaption}
\usepackage{CJKutf8}

\title{Position Paper: \\ Are We Making Progress In Visualization Research?}

\author{Michael Correll\thanks{e-mail: mcorrell@tableau.com}\\ %
        \scriptsize Tableau Research %
}

\teaser{
  \centering
  \begin{subfigure}[t]{0.256\textwidth}
    \centering
     \includegraphics[width=\textwidth]{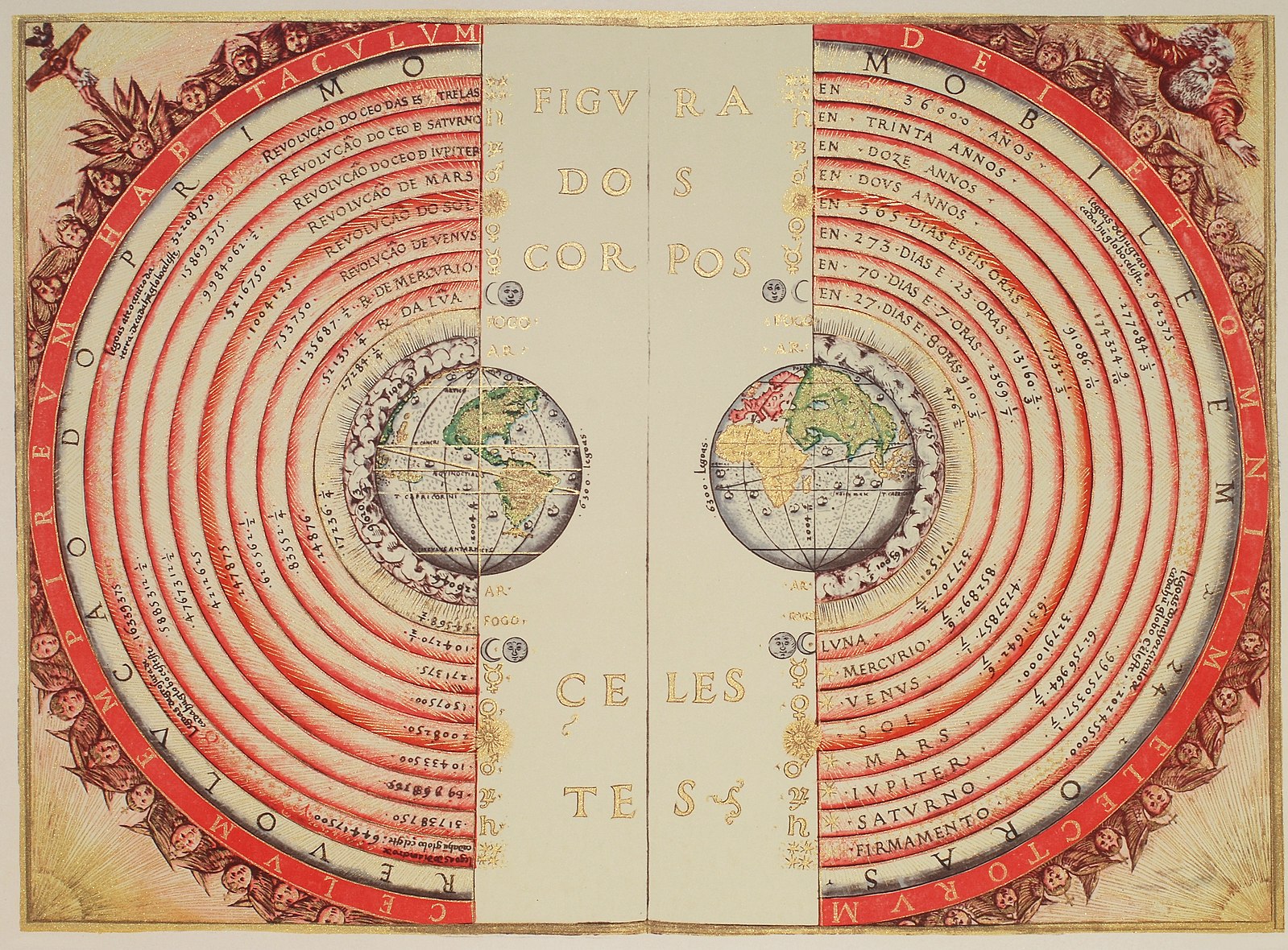}
     \caption{Geocentric model\protect~\cite{geo}.}
     \label{fig:geo}
  \end{subfigure}
  ~
  \begin{subfigure}[t]{0.2\textwidth}
    \centering
     \includegraphics[width=\textwidth]{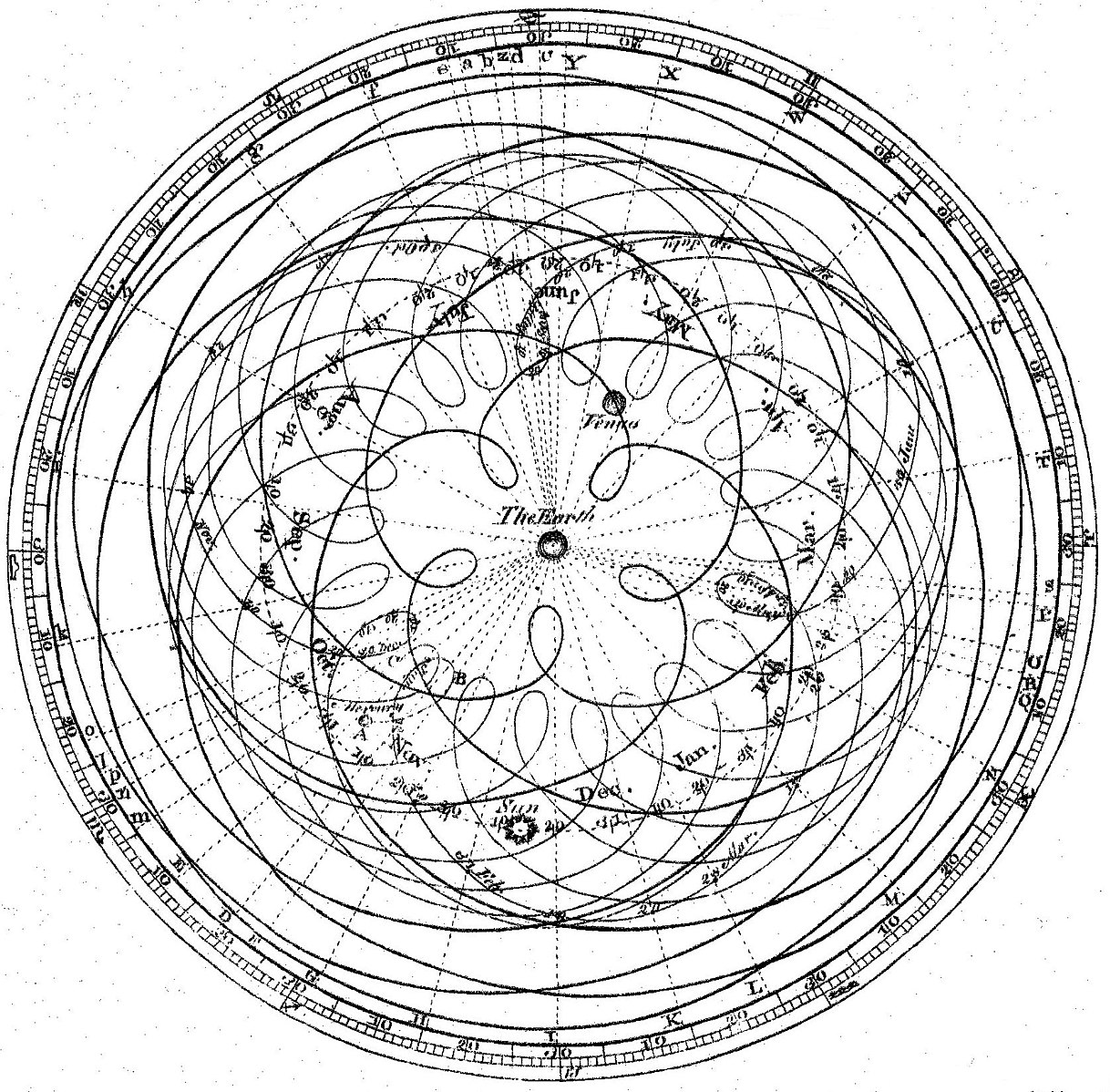}
     \caption{Geocentric model with epicycles\protect~\cite{epi}.}
     \label{fig:epi}
  \end{subfigure}
  ~
  \begin{subfigure}[t]{0.2\textwidth}
    \centering
     \includegraphics[width=\textwidth]{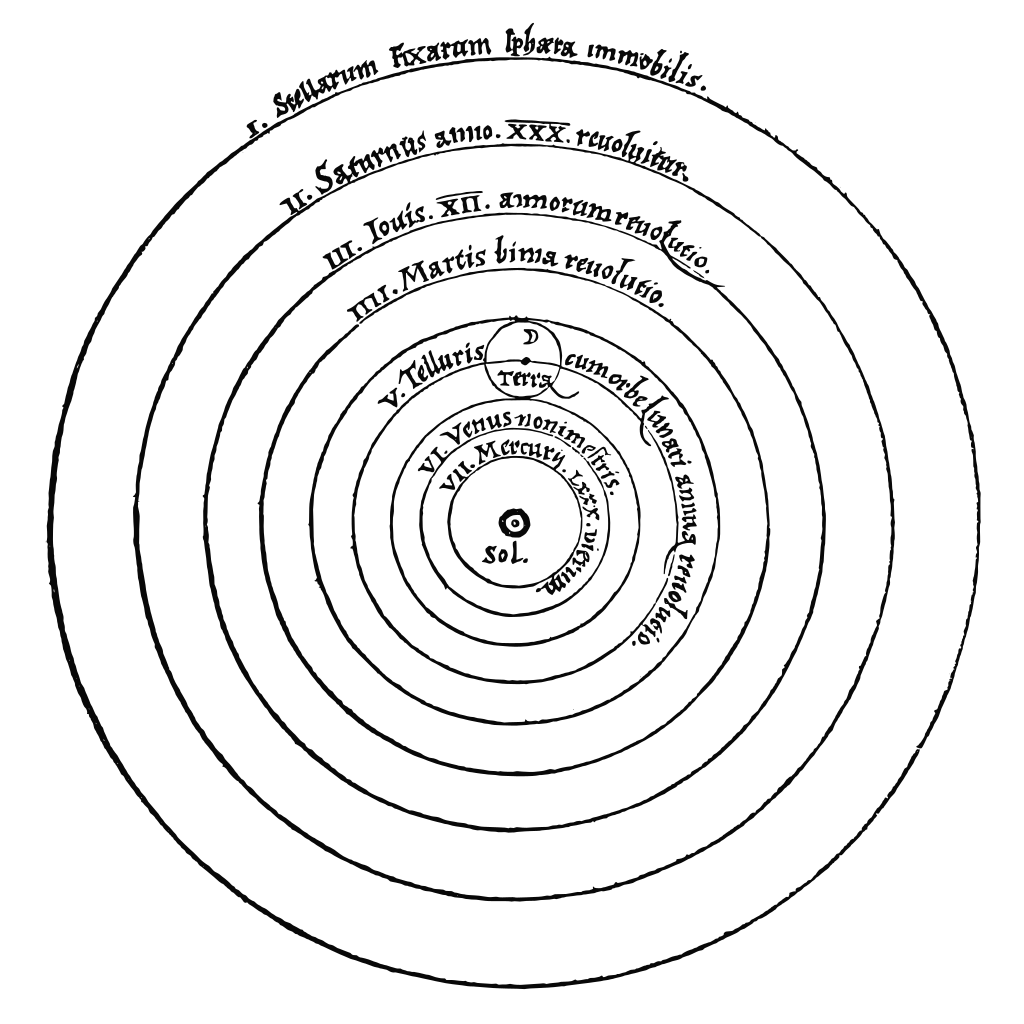}
     \caption{Copernican Heliocentric model\protect~\cite{helio}.}
     \label{fig:helio}
  \end{subfigure}
  \caption{
  Exploring the Bird\protect~\cite{bird2007} model of scientific progress as the accumulation of scientific knowledge: ``an episode in science is progressive when at the end of the episode there is more knowledge than at the beginning.'' The Copernican heliocentric revolution (\autoref{fig:helio}) is progress, even though its initial predictions were not necessarily more numerically accurate than, say, a Ptolemaic geocentric model with sufficient epicycles (\autoref{fig:epi})\protect~\cite{singham2007copernican}. What might an equivalent scientific revolution look like in visualization research, and what are our equivalent of epicycles that add complexity and predictive precision (but not necessarily truth) to our theories? Is such a revolution even possible, for our discipline? Does it matter if it isn't?
  }
  \label{fig:teaser}
}

\abstract{
In this work I use a survey of senior visualization researchers and thinkers to ideate about the notion of \textit{progress} in visualization research: how are we growing as a field, what are we building towards, and are our existing methods sufficient to get us there? My respondents discussed several potential challenges for visualization research in terms of knowledge formation: a lack of \textit{rigor} in the methods used, a lack of \textit{applicability} to actual communities of practice, and a lack of \textit{theoretical structures} that incorporate everything that happens to people and to data both before and after the few seconds when a viewer looks at a value in a chart. Orienting the field around progress (if such a thing is even desirable, which is another point of contention) I believe will require drastic re-conceptions of what the field is, what it values, and how it is taught.
} 

\CCScatlist{
  \CCScatTwelve{Human-centered computing}{Visu\-al\-iza\-tion}{Visualization theory, concepts and paradigms};
  \CCScatTwelve{Human-centered computing}{Visu\-al\-iza\-tion}{Visualization design and evaluation methods}{}
}

\begin{document}


\firstsection{Introduction}

\maketitle



I believe that there will always be plenty of work to do in visualization. There will always be more data to collect and visualize, new or changing environments to design for, and new cohorts of students to train in designing and thinking about visualizations. As individuals, our thinking about visualization and our skills and competencies as researchers also (hopefully) advance over time. So I am confident that there will be, barring catastrophe, a continuing \textit{need} for visualization work, and continuing \textit{individual growth} in service of that work. Similarly, given the continual changes in our environments, tools, and intended users, not to mention the vicissitudes of styles, trends, and resource allocation, I have no doubt that visualization work will \textit{look} different as time goes on, and will likely focus on different \textit{topics} than it does now. So as individuals we will grow, and as a community we will change.

That being said, I emphasize two words from my title: are \textit{we} making \textit{progress}? Where the ``we'' here (rather fuzzily) means the wider community of academic visualization research at large, and ``progress'' (similarly fuzzily, but borrowing many particulars from models of scientific progress from epistemology~\cite{bird2007,sep-scientific-progress}) means ``new, generalizable knowledge.'' Gaining knowledge involves, in turn, strong \textit{justifications} for our beliefs: the avoidance of ``epistemic luck''~\cite{sep-knowledge-analysis} where we blunder into truths by accident, but instead the use of consistent, rigorous, and reliable methods of inquiry. Other fields seem to be able to answer questions about progress with at least the occasional affirmative. For instance, our knowledge of the movement of planets in our solar system increased over the centuries (\autoref{fig:teaser}). But in general, this question is one about \textit{epistemologies}---how do we know what we know? and \textit{practices}---are we doing the things that would help us gain the knowledge we want?

For visualization research, while I am humbled by the effort, ingenuity, and curiosity of visualization researchers, and by no means want to discount their research efforts, I would like to at least introspect on questions of progress, if for no other reason than to be able to build confidence in the longevity and trajectory of the field. What do we know now that a visualization researcher or practitioner would not have known years or decades ago? Similarly, what content (beyond different programming languages or libraries) would we include in a visualization course or textbook now that we would not have included in the past? How did we learn these new things: through controlled experiments, aggregated personal experience, technical innovation, or some other process? Do we practice or teach appropriate or useful methods of inquiry, given the answers to the above? Lastly, and perhaps most pressingly, do we \textit{care} if the answers to any of the above are unsatisfactory, given that we will be able to keep ourselves busy regardless?

In this paper, I analyze the results of a survey of senior members of the academic visualization community on questions of knowledge and rigor in visualization research. These are subjects that require deep thought and can admit many perspectives, especially for fields like visualization that (at the very least aspirationally) incorporate research perspectives from myriad fields like design, statistics, perceptual psychology, and computer science. The limited sample size of my survey, the limited space in this paper, my own limited and situated knowledge, not to mention the observed intractability of these epistemic challenges in fields with longer and more focused pedigrees than visualization, prevent me from claiming to definitively settle these questions. Rather, I use the survey responses and issues raised in prior work as the basis for identifying \textit{struggles} and \textit{opportunities} for epistemic reform or growth as a field. These struggles and opportunities, in turn, suggest ways we could reform or refocus visualization research to promote forward progress (scientific or otherwise) and allow us to build a field that is more robust, more relevant, and more reliable over the coming years and decades.



\section{Related Work}

My questions about progress in a field required me to investigate both of what Niiniluoto~\cite{sep-scientific-progress} calls \textit{backward-looking} and \textit{forward-looking} assessments of scientific progress: ``\textit{If science is viewed as a knowledge-seeking activity, it is natural to define real progress in forward-looking terms: the cognitive aim of science is to know something that is still unknown, and our real progress depends on our distance from this destination. But, as this goal is unknown to us, our estimates or perceptions of progress have to be based on backward-looking evidential considerations.}'' My focus is therefore somewhat bipartite, involving abstract or philosophical assessments of knowledge generation and theory building (which are generally forward-looking) as well as the current or past assessments of rigor, quality, and trustworthiness (which are generally backwards-looking). 

\subsection{Looking Forwards: Building Theory}
\label{sec:theory}

van Wijk~\cite{vanwijk2006views}, lays out a challenge for visualization research that ``\textit{we should ultimately aim at generic results (models, laws) that enable us to understand what goes on and to predict why certain approaches do or do not work.}''  The interrogation of epistemology, the creation or analysis of theory, and the evidentiary backing of visualization research has since been a recurring topic of debate as a community. For instance, a panel at VisWeek 2010 was entitled ``Visualization theory: Putting the pieces together''~\cite{ziemkiewicz2010visualization}, and was followed up by a panel at VisWeek 2011 titled ``Theories of Visualization---Are There Any?''~\cite{demiralp2011theories}. Kosara, both in his role as a panelist on the former panel, and in follow-on work~\cite{kosara2016sand}, questions the strength of the theoretical bases of commonly held assumptions in visualization research. His provocation from his panelist statement is in sync with my own worries~\cite{ziemkiewicz2010visualization}:
\begin{quote}
    \textit{What we need is a new push towards visualization theory: to understand why and how this field works, to find out how it differs from statistics, psychology, computer graphics, etc.; to establish our own basics that we can build on more firmly than the largely ad-hoc approaches today; and to make the case that we do deserve to have our own community, our own conferences and journals, and our own slice of the funding cake.}
\end{quote}
To me, this statement represents an existential challenge to visualization as a unique discipline, rather than an assemblage of particular applied sub-cases of existing disciplines like perceptual psychology or media studies. In some cases, these fields might be pragmatically \textit{better} positioned to attack central research questions than visualization: they might provide easier access to expertise in core methods of inquiry, have longer histories of relevant literature upon which to draw, or be based on more mature pre-existing theoretical or epistemological projects. Even if you reject the premise that visualization is or ought to be a \textit{scientific} discipline but rather a \textit{technical} one (with an emphasis on knowing \textit{how} rather than knowing \textit{that}), practitioners, who may have more training in design, software engineering, or working with customers and clients, might be better positioned to uncover or assess procedural knowledge about visualization.

\subsubsection{Perception as Visualization Theory}
In contrast to this perhaps dismal view, a commonly mentioned theoretical grounding for visualization is the extent to which visualization work is embedded with or draws from perceptual psychology. Ware, in his panel statement~\cite{demiralp2011theories}, makes this lineage explicit: ``\textit{Theory is the means by which experimental results can be generalized and in the case of data visualization in large part this has to be the theory of perception.}'' Rensink~\cite{Rensink2014}, in turn, believes that the ``\textit{prospects for a science of visualization}'' should similar be anchored in research on human perception and cognition. The most salient example of incorporating perception work into visualization design is the ranking of visual channels performed by Cleveland \& McGill~\cite{cleveland1984graphical}. Extending~\cite{kim2018ranking,mackinlay1986}, replicating~\cite{heer2010crowd}, or otherwise revisiting~\cite{bertini2020scatter} this work is a common way of purporting to expand theoretical or epistemic frontiers in visualization research. The specific experimental design and results in Cleveland \& McGill provide benchmarks for exploring new populations of study like crowdworkers~\cite{heer2010crowd},
young children~\cite{panavas2022children}, and even neural nets~\cite{haehn2019cnn}. In turn, empirical results from visualization research can, when collated and structured and reviewed, be legible contributions to the perceptual science literature in return, as in Franconeri et al.~\cite{franconeri2021works}. However, while there are a few evangelists and bridge builders in this interdisciplinary space, these communities are still distinct, and visualization researchers with deep background and expertise in perceptual psychology are (in my subjective opinion), somewhat rare, and the methods and foci of visualization and perceptual psychology researchers overlap but are by no means identical in their epistemologies or goals.


\subsubsection{Procedural Knowledge}
Another common grounding for knowledge in visualization is through of lens of what individual \textit{designs} (and the process of designing them) teach us, such as the popular format of the design study~\cite{sedlmair2012dsm} and the related but distinct concept of action research~\cite{hayes2011action}. Both DS and AR downplay traditional positivist notions of ``generalizability'' or ``replicability'': both forms of design work are about individual populations and situations. However, while acknowledging the subjectivity of the designer and the idiosyncratic nature of the design problem, both aspire to \textit{transferability} as a yardstick for knowledge. This goal of transferability is non-trivial, and may mean that the epistemic value of an individual design study \textit{per se} is relatively low~\cite{correll2020heroic}. We might, for instance, develop transferable taxonomies of design techniques only from collating the work of many hundreds of designs. Or, our knowledge about a single use case might require a additional reflective step where multiple designers mutually reflect on their designs after the fact (Satyanarayan et al.~\cite{satyanarayan2020reflections} comes to mind as an example of one such post-mortem). In any event, while doing design work certainly generates \textit{individual} knowledge and progress (say, in terms of field experience and engineering expertise), it does not \textit{inherently} create useful knowledge for the field without an additional, intentional step of interpretation and legibility. Hayes~\cite{hayes2011action}, building off of Stringer~\cite{stringer2020action}, proposes ``trustworthiness'' as an alternative to generalizability for design work, claiming ``credibility, transferability, dependability, and confirmability'' as necessary components for this goal. Inherent in these desiderata are shared languages, constructs, and design goals: we are not let off the hook from doing theory work or caring about rigor just by virtue of adopting non-positivist ways of knowing.

\subsection{Looking Backwards: Assessing Quality and Rigor}
\label{sec:rel-rigor}
In terms of backwards-looking assessments, there has also been considerable work surveying the heterogeneity and quality of empirical work in visualization. Spyrison et al.\cite{spyrison2021ieee}, for example, finds that \textit{venue}, rather than other factors like open data commitments, is the most consistently top-rated factor in what visualization papers their participants decide to read rather than skim or ignore. Other surveys focus on the empirical methods used in visualization papers. The consistent finding, in my estimation, is \textit{heterogeneity} in methods and practices. Isenberg et al.~\cite{isenberg2013star} found a wide variety of evaluation goals and practices in their overview of VIS papers, with shifting temporal trends pointing to a change in attitudes or cultures around empirical work. Heterogeneity is by no means an inherent flaw, especially for a field with interdisciplinary intentions and participants: for Wall et al.~\cite{wall2022vishiker}, the ``opposing lineages'' of visualization research are the main driver of both conflict (in dispositions towards study design) but also opportunity (to create a ``breeding ground of innovation'' for answering new types of questions in new ways). Meyer and Dykes~\cite{meyers2020}, in their assessment of rigor in qualitative work, likewise point to the benefits of multiple perspectives in empiricism, and lay out a case that multiple ways of knowing can co-exist while still building towards accepted standards of quality, rigor, or credibility. Conversely, \textit{homogeneity} in methods of inquiry can be damaging: Hullman et al.~\cite{hullman2019uncertainty} felt that the relative uniformity of analytical ``paths'' taken by their assessed corpus of studies of uncertainty visualization created potential threats to ecological validity, for instance by focusing on accuracy and efficiency of extracting values rather than performance at wider sets of actual decision tasks.

However, heterogeneity does introduce challenges by making standardization, re-use, and collation of results difficult. The many ways that ``tasks'' are defined or tested in visualization studies in Pandey et al.~ \cite{pandey2020tasks} was mentioned as a field-wide obstacle, despite the near-ubiquity of task analyses in the design and evaluation of academic visualization work. In critiquing evaluations of visualization in prior work~\cite{correll2020heroic}, I likewise decry the lack of interoperability in visualization papers: ``\textit{at the very least it seems rude to future generations of researchers to make them have to pick through the rubble of our current practices to find the few apples-to-apples comparisons they can salvage.}'' Yi's panel position statement~\cite{ziemkiewicz2010visualization} makes this point more directly and in terms closer to the goals of this paper:

\begin{quote}
    \textit{I believe that in order to build useful theories of information visualization, we should first collect reliable and comparable empirical evidence of how information visualization is used and work, so that researchers can identify patterns, propose hypotheses and theories, and test them. However, in the field of information visualization, we have not established a proper culture to collect comparable data, yet.}
\end{quote}

These issues of inter-operability or legibility are merely prerequisites in order to even evaluate the existence of other potential threats to rigor, reliability, or credibility. Even when our results are comparable, our empirical methods may fall prey to many of the same issues that resulted in replication crises or other crises of confidence in other fields~\cite{kosara2018,cockburn2020}. Already, claims about the persuasiveness of visualizations~\cite{dragicevic2018} or the impact of anthromorphic visualizations in eliciting empathy~\cite{morais2021}, or more foundational bits of visualization folklore like the importance of ``banking to 45''~\cite{talbot2012banking} have been questioned by re-analysis and additional scrutiny.


\section{Methods}

I designed a survey meant to assess attitudes around progress, rigor, and knowledge in visualization research, and in particular in research presented at IEEE VIS. I focused on open-ended, wide-ranging questions. I also focused on questions that asked for longitudinal or overarching perspectives. These choices no doubt complicated my analysis and reduced both the sampling pool and the completion rate, but I judged the potential to provoke richer and wider-ranging responses worth the risk and effort. 

I hosted the survey on the SurveyMonkey platform and sent out an email invitation to participate to an initial list of senior visualization researchers and thinkers. This initial list was chosen by me based on prior publications around epistemological or methodological issues in visualization, internal reputation for interest in these issues, and through combing the organizing committees of VIS workshops or panels connected to these issues.

To suppress (but by no means eliminate) the selection biases inherent in such a list, I iterated on and refined both the question list and the initial participant invitation list based on feedback from three colleagues. In addition, I employed snowball sampling through the use of a final question, ``Who is someone else you think I should send this survey to?'' In all, I sent the survey to 47 participants, of whom 14 completed the survey. While I used my initial participant list to \textit{invite} participants, in the interest of collecting minimal personal data and reducing overhead I did not place any access management or tracking on the survey itself: as such, I cannot discount (but must admit I am not too worried by) the possibility that people not on my list took the survey (for instance, by being emailed the survey link by a colleague).

This final invitation list, a PDF version of the online survey, along with lightly edited and anonymized (when requested) responses are available at \url{https://osf.io/nzpka/}.

\subsection{Questions}

In addition to a set of demographics questions and space for comments or reflections on the survey itself, I focused on three sets of questions, organized around topics of priorities, knowledge, and rigor. I include the actual question text within the descriptions below:

\textbf{Magic Wand Questions}: Questions assuming the respondent has arbitrary power to shift disciplinary foci or reward structures, meant to assess both priorities and worries about progression in visualization. Namely:
\begin{enumerate}[nosep]
\itemsep0em
    \item You can wave a magic wand and create one new \textit{requirement} for research papers in visualization. What is it?
    \item You can wave a magic wand and create one \textit{new reward or award} for research papers in visualization. What is it?
    \item You can wave a magic wand and make the academic visualization research community spend \textit{more time} on one problem or one area of research. What is it?
    \item You can wave a magic wand and make the academic visualization research community spend less time on one problem or one area of research. What is it?
\end{enumerate}

\textbf{What Have We Learned? Questions}: Questions about individual or collective knowledge over time in visualization, meant to assess changes in knowledge but also, as a follow up, to solicit \textit{ways of knowing} in visualization. Namely:
\begin{enumerate}[nosep]
\itemsep0em
    \item What’s something \textit{you} know about visualization that you didn’t know when you started in the field? How did you learn it?
    \item What’s something in visualization that \textit{you} have been wrong about or otherwise had to reconsider in light of new evidence? How did you accept you were wrong?
    \item What’s something that \textit{the field} generally knows about visualization that we didn’t know 10-20 years ago? How did the field learn it?
    \item What’s something in visualization that \textit{the field} has been generally wrong about or otherwise had to reconsider in light of new evidence? How did the field accept that it was wrong?
\end{enumerate}

\textbf{Rigor and Methods Questions}: Questions about (usually comparative) perceived rigor in visualization work, meant to assess the perceived quality of our epistemic tools. Namely:
\begin{enumerate}[nosep]
\itemsep0em
    \item If you read or write \textit{papers} for other fields (e.g., psychology, graphics, design), are papers in visualization (or IEEE VIS specifically) more or less rigorous compared to papers in those other fields? How, and in what ways?
    \item If you read or write \textit{reviews} for other fields (e.g., psychology, graphics, design), are reviews and reviewing of visualization papers (or IEEE VIS papers specifically) more or less rigorous compared to reviewing in those other fields? How, and in what ways?
    \item If you design, review, or perform research that employs \textit{quantitative methods}, are the quantitative methods commonly employed in visualization (or IEEE VIS specifically) more or less rigorous compared to other fields? How, and in what ways?
    \item If you design, review, or perform research that employs \textit{qualitative methods}, are the qualitative methods in visualization (or IEEE VIS specifically) more or less rigorous compared to other fields? How, and in what ways?
\end{enumerate}

\subsubsection{Survey Limitations}
I did not intend for these questions to be comprehensive, even putting aside the difficulty of impossibility of capturing epistemic attitudes \textit{in toto}. For one, I wanted to reduce the burden on the survey participants (especially given my intended subject pool, who I assumed to be under considerable competing pressures for their time and attention) by employing a smaller set of open-ended answers that afford variable response length and engagement. Also, even these current questions, requiring as they do ideation, theorizing, and reflection, blur the line between \textit{participant} and \textit{co-researcher}: asking for additional intellectual expert labor without offering more concrete rewards for participation (such as co-authorship or consulting fees) was personally uncomfortable for me.

Another salient limitation was a lack of core \textit{definitional} questions (e.g., ``Define `progress' in your terms''), which I intentionally excluded as both too abstract but also difficult to answer without significant theorizing; in retrospect, these definitions were sufficiently core to my research questions that their omission curtailed my project.  Other significant questions were relegated to followups (like ``how did you learn it?'') that were either ignored by respondents or dealt with in less detail in favor of the primary question.

\subsection{Participants}

14 participants completed the survey. For each participant, I donated \$10 USD to a preferred charity from a list of three I supplied.

In keeping with my intended participant pool of senior or established visualization researchers and thinkers, participants reported a mean of 17.1 (SD = 6.7) years of experience in the visualization community. 11 participants reported a role as an academic or professor. 11 participants regularly taught visualization or visualization courses or workshops, 2 did so occasionally, and only one did not self-report as teaching in visualization. 12 self-reported regularly submitting work to academic conferences or co-located events with a focus on visualization work. I intentionally did not collect demographic information such as age or gender, but allowed participants to report any additional demographic information of relevance.

\subsubsection{Attribution}

I offered participants the option to have their responses attributed to a self-reported name or initials, role or title, by a numerical participant ID, or not attributed directly at all but only used as part of aggregate thematic analysis. All participants who completed the survey agreed to direct attribution, but there was heterogeneity in preferences beyond this point. Eight participants wanted their statements attributed to their role: P1, P2, P5, P8, P9, and P14 refer to themselves as academics, P7 as a senior visualization professor and P10 as a data visualization engineer. Three participants asked to have their statements attributed to a name or set of initials: P4 is HLP, P11 is Steve Haroz, and P12 is Enrico Bertini. I followed up with Drs. Haroz and Bertini through other channels to confirm that the statements attributed to their names were in fact made by them. The remaining three participants, P3, P6, and P13, asked to be identified only by number. To avoid stilted language, I use participant numbers exclusively in the remainder of this paper. 

\subsection{Analysis}
I read the full transcripts of all respondents and identified repeating themes or topics of personal interest within my three categories of questions. The tags I used to flag these themes, and brief explanations of their meaning, are included in my osf repository, but generally correspond to rough labels of \textit{topics} of responses, rather than the \textit{valence} of these responses. E.g., the ``Theory'' tag referred to both responses asking for an increased focus on theory-building in visualization, but also those suggesting that theory-building was not necessary or counter-productive. Given the small sample size and the highly idiosyncratic nature of solo-coding, I endeavor where possible to let participants speak in their own words rather than reporting on the frequency of recurring themes.


\section{Results}
This section is organized around my emergent themes. I note that my questions included specific calls to critique as well as comparisons to ideal states. As such, many of the responses center persistent or recurring \textit{struggles}: obstacles to the growth of the visualization field, or areas where there are calls for improvement. I discuss \textit{opportunities} in \autoref{sec:hdi} and \textit{counter-narratives} in \autoref{sec:mu}.

\subsection{Methodological Rigor}
\label{sec:rigor}
Rigor is a core component of judging scientific progress. We do not have direct access to the truth, and so must instead use the strength of our \textit{methods} to gauge whether or not our conclusions are well-founded. I should note that the questions I asked were not exhaustive of the methods employed in visualization work (missing, for instance, are algorithmic analyses, literature reviews, or distinctions between the layers of analysis in common views of empiricism in visualization such as Munzner's nested model~\cite{munzner2009nested}). I would also take care to point out that deficiencies in one area may or may not generalize to statements about the state of the field as a whole. In the words of P14:


\begin{quote}
    \textit{To me, a big difference of the VIS community is its breadth, including design, engineering, and evaluation concerns. I think VIS often has a more rigorous *integrated* perspective, but less rigorous around specific sub-areas or tasks. True interdisciplinary (transdisciplinary?) work is hard.}
\end{quote}

\subsubsection{Quantitative Rigor}
Impressions of the rigor of quantitative work in visualization were mixed. Some respondents were positive (or at least leaning positive): P12, for instance, said ``\textit{I am not sure but my general sense is that vis is quite rigorous compared to other areas but often not rigorous enough}.'' Likewise, P7 claimed: ``\textit{VIS has actually led the way in many quantitative evaluation methods, so I would say they are often more rigorous (at least in the last 10 years) than for HCI papers.}''

However, other respondents were quite negative about the state of quantitative work in VIS. Two responses even express doubt over the benefits of much quantitative work in visualization \textit{at all}. P11 stated that he was wrong about or otherwise had to reconsider  ``\textit{that behavioral research by the visualization community has any value to the visualization community. In over a decade, I haven't seen any}'' and wished for a requirement that VIS papers authors be ``\textit{competent in the methods they use.}'' P3 was also skeptical of current quantitative approaches in visualization:
\begin{quote}
\textit{Dataviz papers often fail to specify a model for behavior and instead jump right into the results. Dataviz papers also claim to meet a high quantitative bar, but the sample sizes are woefully small. It seems that the dataviz research field struggles with wanting to be a quantitative science like economics, but the research is actually closer to a qualitative science like behavioral psychology or sociology.}
\end{quote}
Most damning for the premise of this paper, P12 was similarly unconvinced of the ability of current studies in visualization to generate new, generalizable knowledge:
\begin{quote}
    \textit{[I now know t]hat scientific ``truths'' are way harder to get than we believe. I am way more skeptical about studies than I used to be. I am increasingly worried they are all very limited in a fundamental way.}
\end{quote}

While less negative, other respondents felt that quantitative work at VIS compares unfavorably to other fields. P1 summed up this position by saying ``\textit{[quantitative methods] are often less rigorous [in VIS]. Deep statistical knowledge is required which is often not part of the training of vis experts.}'' Psychology as a discipline was a recurring yardstick. P6 in particular suggests that visualization work often comes up short in comparison to psychology: 
\begin{quote}
\textit{Compared to journal articles in psychology, VIS and CHI papers receive far less scrutiny, fewer rounds of editing, less competent/engaging peer review, and have much shorter project timelines. Graduate students in CS-adjacent fields tend not to receive high-quality training on research methods, and we are incentivized to rush projects to meet conference submission deadlines. The result is not so much that VIS and CHI papers are less rigorous on the whole but more so that a lot of mediocre work gets submitted and published.}
\end{quote}

P14 echoes this sentiment, but acknowledges variability:
\begin{quote}
   \textit{Compared to psychology, I find visualization papers are typically less rigorous in terms of experimental design and methods, everything from power analysis to construct validity to statistical analysis of results. [...Quantitative methods in VIS are l]ess rigorous on average compared to psychology, though certainly improved versus 20 years ago. [...] The best work can be quite rigorous, but less rigorous work can make it through the review process. I believe reviewer background and education remains a key issue here.}
\end{quote}

P6 goes even further with claims of variability:
\begin{quote}
    \textit{It really depends on the authors. I would say that vis research has both some of the best and the worst quantitative work that I've seen. Some vis researchers apply quantitative methods with a level of technical sophistication, polish, and multidisciplinary vision that is seldom achieved in other disciplines. Some vis research apply experimental design and statistics in rote and mistaken ways because they simply don't know any better. I don't fault them because it's an issue with training.}
\end{quote}

\subsubsection{Qualitative Rigor}
While there were some defenders of the quantitative rigor of VIS papers, especially compared to other areas of the HCI ecosystem, there were only two respondents who provided positive comparative pictures of qualitative work, neither of which were full-throated. P8 claimed ``\textit{We're probably better than most fields at qual work.}'' whereas P12 responded ``\textit{I am not sure. I'd say most of the best [qualitative] researchers are pretty rigorous.}''

The other responses were generally more negative about qualitative rigor in VIS. A common critique of qualitative work was a lack of structure, especially compared to other fields. P1 claimed ``\textit{Structured and solid methods for qualitative analysis are often not applied and rather ad-hoc reporting is being done}'' and P5 claimed that qualitative work was ``\textit{less rigorous, if only because of the reason that these approaches are new to the community so many are learning by doing.}''

As with psychology as a benchmark for quantitative methods above, here sociology and anthropology were frequent points of comparison. For instance, P13 stated:
\begin{quote}
\textit{I think [qualitative methods are] less sophisticated in VIS than in other disciplines. In sociology for example, folks are often up front about their approach and underlying philosophical position: here I take phenomenological perspective ... etc. I think there is a naive assumption in VIS that ``looking at the data'' is enough.}
\end{quote}

However, even in critique, respondents were appreciative of the potential benefits of qualitative work. P14 stated:
\begin{quote}
\textit{Both [VIS and CHI] are often much less rigorous than fields like anthropology. However, I also think the aims are different. For an extreme contrast: long-term ethnographic studies informing/critiquing social theories are a rather different enterprise than semi-structured interviews conducted to inform a design activity.}
\end{quote}

And P6 echoed the above sentiment nearly exactly:
\begin{quote}
\textit{...[Qualitative] methods in tech are utilitarian and tend to cut corners that might offend an anthropologist. That being said, they often teach us things that are valuable to know and otherwise hard to study.}
\end{quote}

I should note that the negativity or lack of certainty might be a sampling bias due to the relative newness of qualitative work in the field, newness in exposure to concepts of qualitative rigor~\cite{meyers2020}, or a lack of expertise in my participant pool: P3 responded to my question about quantitative rigor with ``\textit{I can't speak to this question with any real authority}'' and four respondents either left the question blank or responded with ``\textit{N/A.}''

\subsection{Utility and External Validity}
\label{sec:validity}
A recurring concern with visualization as a discipline is a potential lack of focus on the applicability of results, and a perceived lack of concern for potential real-world uses of visualization research. I should note that these are not necessarily critiques of progress \textit{per se}, but of \textit{useful} progress. That is, the researchers might be learning new things, but these things are not \textit{useful}, either in the sense of utility for moving the field forward, or in the sense of not producing knowledge that people outside of academia would care about.

\subsubsection{Bespoke Tools}
A frequent claimed research contribution in VIS is the building of new tools, often for a small audience of domain collaborators. Respondents were often skeptical of the value of this work. P4, P12, and P3 suggested that the community spend less time on ``\textit{One-off designs}'', ``\textit{Design studies}'', and ``\textit{Tools. We don't need more research on one-off tools}'' respectively. P14 echoed this sentiment but provided more detail, suggested that the community spend less focus on ``\textit{bespoke visual analysis systems (typically consisting of multiple coordinated views backed by some analytics algorithm(s)) lacking sustained use, maintenance, or larger lessons learned.}''

P4 was particularly critical of application work in visualization, both in how it is written up in papers but also how it is reviewed compared to other fields:
\begin{quote}
    \textit{There are also a lot more papers of the form ``there is a specific analysis problem and I designed a specific solution/system/design/etc for this problem''. [...] There is less focus on importance of a paper's idea, and more focus on novelty. In [non-VIS] systems papers, there's an implicit measure of how novel a solution is \textit{and} how important/impactful it is as well.}
\end{quote}

Despite this skepticism, systems and applications work was acknowledged as difficult, important, and an area where visualization was seen as making clear progress. For instance, P14 claimed ``\textit{I think we have made great strides in how to design and engineer both languages and systems for visualization.}'' P8 claims that ``\textit{we've seen a lot of progress on the technical front (better libraries and systems)}'' but, in their response to another question, thinks that we should introduce new awards so that ``\textit{[people will] be incentivized to build open source libraries and tools beyond just prototypes.}'' 

\subsubsection{Escape}
\label{sec:escape}
I have in the past used ``escape'' as a term for a potential disciplinary failure where effort or discussion is ``moving so far away from the political or technological realities on the ground that we cease to have any impact whatsoever.''~\cite{correll2021critihype}. Some respondents suggested that the field is at least on a metaphorical escape velocity. For instance, P2 claimed ``\textit{the field still does not accept that most visualizations published in viz are unusable for most people}'' whereas P1 wanted the community to spend more time on ``\textit{increasing the visibility and appreciation of visualization research outside the core vis community}.'' While acknowledging a need for diverse community foci, P13 also asked if we ``\textit{might spend less time in the lab and more time talking to people}.'' P6, attacking a similar issue, found the lack of design implications a recurring problem for both study- and application-based visualization work:
\begin{quote}
   \textit{Too often I see papers without a good concept, e.g., psych studies on basic research questions without real implications for practice, or system building papers with no underlying theory or implication beyond the target domain.}
\end{quote}

There were several calls to remedy a perceived lack of real-world relevance of visualization research. P4 stated that papers should be required to have ``\textit{less rhetoric, more operationalization.}'' A more detailed paper requirement was from P10, who called for  ``\textit{evidence that the kind of work being research[ed] has been used in practical applications outside academia. This could include showing evidence of prior approaches to this topic being seen `in the wild''}' and, beyond this requirement, creating a ``\textit{test of time award that demonstrates a technique was adopted and productionalized in the real world.}'' This respondent also suggested that the visualization community should spend less time on the ``\textit{readability/comprehensibility of data visualization by students and mechanical turk participants rather than invested domain experts}'', and in fact critiqued \textit{the survey itself} for failing to consider practitioner perspectives: ``\textit{there seems to be no thought of engagement with the design community just the technical community.}''

Another point of contention was the inadequacy of current theoretical structures (such as the ranking of visual channels in terms of effectiveness) as guiding or useful theories for visualization design. P8 claimed ``\textit{We now know that it isn't always about ``precision'' of a visual encoding}'' while P14 claimed ``\textit{I think the field has been (as a whole) rather unquestioning of quantitative proportional judgments as a sufficient proxy for perceptual effectiveness}.'' And P4 claimed that ``\textit{We know basically nothing about graphical perception.}'' P12 was even more explicit on this point:
\begin{quote}
 \textit{I have been wrong in believing one could produce effective visualizations using the ranking of visual variables as the main guidance. I accepted it by being increasingly exposed to my student's criticism and questions. After a few years I was forced to admit it does not make much sense as a guiding theory for vis.}
\end{quote}

\subsection{Human-Data Interaction}
\label{sec:hdi}
While the prior sections deal with what I consider \textit{struggles} in visualization research (areas where we are currently engaged but have a potential need or capacity to improve), the issues raised in this section I view as more of an \textit{opportunity}: an area where the field has the potential to make formative changes and do qualitatively different kinds of research. In particular, multiple respondents wanted the field to focus on, in the words of P5, ``\textit{the human ways that visual analysis is done}'' or, from P2, ``\textit{insight into cognitive foundations of higher-level information visualization}.'' I borrow the term Human-Data Interaction (HDI) to describe this gestalt notion of centering the individual or the organization within the process of data analysis, and moving from either statistical or perceptual models of understanding visualization and to meta-cognitive or sociological perspectives. The IEEE VIS 2021 HDI workshop~\cite{hdi} defines the project thusly: ``\textit{[t]he emerging area of human-data interaction (HDI) encompasses all aspects where humans touch and engage with data, widening the scope beyond traditional visualization and visual analytics to consider the breadth of how people think with and use data.}'' For instance, rather than studying atomic task efficiency, P3 suggests that the field could:
\begin{quote}
    \textit{[Spend more time learning h]ow practitioners actually create data visualizations and meet the needs of their users, readers, or audience members [and do m]ore work on successful organizations and teams creating effective visualizations. What are effective team structures, work processes, or data workflows?}
\end{quote}

P6 in particular welcomed this shift as a result of a feeling that we have moved past or otherwise saturated other forms of inquiry: ``\textit{[h]ow to structure people's thinking around vis seems increasingly more interesting as a problem than comparing visual encodings.}'' P6 expanded on this sentiment in another response:
\begin{quote}
    \textit{[I've learned that v]isualization is a social object as much as a computational one. I think we've mostly learned this because the field has become increasingly interdisciplinary and because people have become somewhat bored with all but the most innovative technical work.}
\end{quote}

This focus on HDI, and consideration of larger units of study (the entire data pipeline rather than the ``final'' visualization, the sociotechnical milieu of data analysis rather than a single data task), requires, to my eye, new methodologies, lenses, and even time frames of visualization work.





\begin{CJK*}{UTF8}{min}
\subsection{無}
\label{sec:mu}
In a popular Zen koan, a monk asks J\={o}sh\={u}, ``Does a dog have Buddha nature?'' and is told, in response, ``無'' (mu). Hofstadter~\cite{hofstadter1979godel} and other Western commenters often translate 無{} in this koan as being a negative response that also has connotations of unasking or otherwise negating the premise of the question itself. 

I used the tag ``mu'' for responses that questioned or negated the premise of my questions, or indeed the entire framing of my research project. Some of these 無{} responses were relatively low-level: for instance, P11 responded to the question about what ``the field generally knows'' with ``\textit{I'm not sure how or if `the field' can know things}.'' Similarly, multiple participants objected to the premise of my question around what ``reward or award'' they would create: P9 stated ``\textit{Why do we need another reward/award?}'' and P13 maintained ``\textit{I am skeptical about these awards. They result in narrowing.}''

But while some of these 無{} responses point more to potential deficiencies in my survey design, others raised important questions about the very \textit{desirability} of the project laid out in my paper title, either in terms of whether it is useful to think of visualization as a field that does or can care about scientific progress, or whether or not calls for narrow definitions or rigor or progress would negatively impact our aspirations to be a diverse and hetereogeneous field or otherwise prevent us from doing useful work. van Wijk~\cite{demiralp2011theories} echoed these sentiments in his panel statement: ``\textit{The discipline of Visualization is not a science, it's technology. Our aim is not to develop theories about the world or the universe; we try to develop methods and techniques that enable people to do their job more effectively, efficiently and with greater satisfaction.}''

While blunt, the statements by P9 that visualization research should ``\textit{spend less time erecting walls and trying to define what is and isn't visualization research}'' and P13 that visualization research should similarly ``\textit{SPEND LESS TIME TRYING TO COME UP WITH A UNIVERSAL THEORY OF EVERYTHING AS THOUGH THIS WAS PARTICLE PHYSICS}'' [all caps in the original] illustrate these point of views concisely. Respondents have noted that their thinking has evolved on this issue. P5 said they ``\textit{[were wrong to think] that problem-driven vis research, system research, or vis design research (basically everything except cogsci studies) can be described, conducted, and analyzed effectively from a scientific/positivist perspective.}''

This quote by P13, although extensive, I believe is worth reproducing in its entirety in this context as a \textit{cri de c{\oe}ur} over the potential benefits of assessing rigor, but also the potential hazards of choosing definitions of progress that are too narrow:
\quote{
\textit{It would be good to get some sense of what people think rigor is and how (why, whether) they think it's important. I doubt people think about this much to be honest, I think it's is kinda taken for granted, but I would like to get them doing so much more. Depending on what rigor is and what you are trying to do, it may or may not be important. we really ought to know more about what we are looking for and what we expect in great quality work. The excitement about and strength of VIS is that it is so varied and diverse. Sometimes, making something (I nearly said *just* making something there) is informative and generative and persuasive and results in knowledge. And then someone comes along and asks for a user study because they have a limited view on how we can generate useful and reliable knowledge. If we are not open to different epistemological possibilities to fill the vast colourful space of VIS and try to limit these in some kind of dumbass quality control exercise then we will end up with a very dull and diluted discipline. I don't think people see this. Hopefully this exercise will prove me wrong.}
}
\end{CJK*}

\section{Discussion}
The responses to my survey, most saliently, should disabuse anyone of any notion that there is a single shared view of what visualization research is, was, or ought to be. This lack of a shared epistemic project is perhaps bad news for someone hoping to propose a single set of field-wide requirements or syllabi for conducting rigorous, credible, or useful visualization research, but was acknowledged by many respondents as a strength, rather than weakness, of the field. An interdisciplinary field, after all, requires a multitude of perspectives. The answer to ``are we making progress in visualization'' is therefore, predictably, ``it depends.''

All that being said, I do not think the existence of multiple, occasionally conflicting perspectives renders the prospects of improving rigor or setting more impactful research trajectories fruitless. At the risk of offending P4, who stated ``\textit{It feels like there's a lot more speculation and flash in VIS papers. Anything seems to go in a discussion section}'', I will use this section to speculate about how we might institute reforms or goals that, at least for particular subsets of visualization research, might move things forward. These suggestions are to some sense oppositional, at least strategically, but I think at the tactical level would share many similarities.

\subsection{Address Rigor}
While there were many comments speaking to comparative issues of rigor in visualization work, there are several competing obstacles for anyone seeking to address these issues of rigor or research quality. The first is wide disagreement about the degree and scope of the issue: other respondents were relatively positive about the degree of rigor in VIS, especially compared to closely related communities in HCI, for instance. The second is, even if the problem(s) are shown to be severe, we would still need to stir people out of complacency. The last obstacle is training and building internal expertise: I don't think even the most negative respondents in my survey would attribute methodological weaknesses to some inherent idiocy among visualization researchers, but rather a lack of training or, perhaps more to the point, the existence of many competing areas where visualization researchers need training or could otherwise prioritize: visual design, software engineering, experience in applied domains, research methods, and many many other areas where expertise is required or expected to make an impact.

Therefore, a successful methodological reform movement in visualization would, for me, have the following characteristics:
\begin{enumerate}
    \item \textbf{Support for the many ways of knowing in visualization and types of visualization work.} There are many many ways to make a contribution to visualization, not all of which have inter-operable conceptions of rigor (or even a particular interest in the concept). Several respondents made a particular point of the interdisciplinary and heterogeneous nature of visualization research as being both a strength of the field as well as a complicating factor for generating universal requirements or standards of rigor. That being said, there are still opportunities (say, by leaning on the  IEEE VIS ``Area Model''~\footnote{http://ieeevis.org/year/2022/info/call-participation/area-model})  to create explicit per-area requirements or expectations to promote both \textit{global} diversity in methods while supporting \textit{local} rigor. For instance, having differing requirements for differing types of contributions. As an example, an ``open data'' requirement would be necessarily different for a graphical perception paper (where both the experimental stimuli and the analyses should be shared), compared to a paper promoting a new rendering algorithm (where perhaps only the source code or even just pseudo-code would be necessary).
    \item \textbf{Longitudinal interventions.} If current methodological weaknesses are due to a lack of training, then it suggests that, say, immediately and universally creating higher bars for paper rigor would succeed mostly in locking out potentially large portions of the current field. While the field can (and does) change norms quickly, and some short-term interventions seem quite feasible to me (for instance both P1 and P12 suggested open data, or at least access to working demos, should be a requirement for VIS papers), others might require longer-term planning. I am thinking here of new or expanded textbooks, new curricula, and organizational experiments (some of which might very well fail!) in how rigor is measured or rewarded in reviewing or publication, which are efforts that might take years or (academic) generations to produce conclusive results.
    \item \textbf{Rewards as well as punishments.} There is a reason why positive reinforcement has pride of place in behavioral research. Under the assumption of systematic issues in methodological rigor (rather than mere variability in quality), solutions like rejecting more papers are merely a form of punishment. To make matters worse, since these sentiments about rigor are not universally shared even within the set of respondents in this paper, new and unilateral higher bars of rigor are also punishments on a \textit{random schedule}, influenced by the lottery of individual reviewers rather than the predictable outcome of work of insufficient quality. I therefore argue that rewards and other ways of reinforcing positive behavior would likely increase the impact of any reform effort more than punishments alone. I, personally, am not so anhedonic that I immune to the morale-boosting impacts of a certificate or a footnote on a CV for doing rigorous work.
\end{enumerate}

\subsection{Lean on Other Fields}
\label{sec:lean}
Another option for progress is to lean more heavily on the ``source'' disciplines that make up visualization. The word ``lean'' here is intentionally under-specified, but in this umbrella I include a number of potential activities. For instance, if we think that, say, perceptual psychology is a core part of building theory in visualization (I choose this example only because it is often proposed for this niche; mentally insert ``social science'' or ``statistics'' or ``software engineering'' etc., as per your disciplinary proclivities), then we could require students to take perceptual psychology courses as part of a standard visualization curriculum, or attempt to publish or make our experimental work legible and acceptable to the perceptual psychology community \textit{first}, with the work in VIS focused more on the pure application of these empirical findings to visualization design. 

This intervention could go so far as to reduce or eliminate visualization as a ``home'' discipline entirely. For instance, if one is proposing a new visualization application for biomedical data, the authors would focus on first publishing work on the tool (or the novel results that the tool enabled) in biomedical journals, with subsequent papers in VIS appearing only afterwards and reporting on a much narrower scope of contributions. This would both shore up the credibility of any claims of utility or rigor in the tool (since it had already been vetted, however imperfectly, by the applied domain) and allow VIS paper to focus more on the visualization-specific aspects of the work, but would still ``reward'' thorough and useful (but perhaps not particularly novel or field-changing) design work. This refocusing could also involve alternative ways of doing work, such as the ``design study lite''~\cite{syeda2020lite} methodology that values shorter-term and more pedagogically-focused design interventions rather than a full design study~\cite{sedlmair2012dsm} that is teleologically oriented towards writing up results for an academic audience.

I must admit that I find aspects of the above solution personally unpalatable: I'm not sure how it applies to domain-agnostic systems or design work, I think it would reduce feelings of belonging or the professional benefits of centralization, and might encourage playing (more) games with publications strategies. A less radical but still potentially useful effort would be to continue to strengthen our connections with our component disciplines. I note that these interdisciplinary connections (and individual transdisciplinary connectors) already exist in the community, but often in the periphery of the main conference events (for instance, there was a ``VisPsych'' workshop at IEEE VIS 2020~\cite{vispsych}, following up several years of meetups). This strengthening could include outreach efforts to other conferences or disciplines, recruiting from those disciplines for reviewing, speaking, or teaching, or even simply making an effort to periodically ``report out'' to other audiences in ways that are legible and useful without having to be embedded in academic visualization jargon or perspectives. Getting good at this sort of connection-building would also have knock-on effects for other areas where visualization was seen as falling short by my respondents, such as providing utility to visualization practicioners (\autoref{sec:escape}).

\subsection{Find Our Own Voice}
The last avenue of reform I propose is also perhaps the most difficult to conceptualize, let alone operationalize: building a \textit{strong and shared epistemic and theoretical foundation for visualization work}. This would mean to seriously take up the project proposed by Kosara and others in \autoref{sec:theory} and determine what quintessentially sets apart visualization from being just ``applied perceptual psychology'' or ``applied computer graphics'' or ``applied design.'' I believe there are unique intersections and unique perspectives that arise from thinking deeply about \textit{data}, about how the data are \textit{represented}, how they are \textit{perceived}, and, finally, how people use those representations to \textit{think}. Visualization in this lens is not a hodgepodge of differing perspectives but a cross-cutting effort to learn something new about the world given the unique expertise and knowledge of visualization researchers. HDI and other lenses (such as perspectives (re-)thinking visualization in information theoretic~\cite{chen2010information}, inferential~\cite{Hullman2021Designing}, or consequentialist terms~\cite{vanwijk2005value}) I feel (and at least some others, see \autoref{sec:hdi}) are promising avenues for considering larger components of information design and understanding. There are many potential candidates for constructing a unifying theory of visualization, with hypotheses to test or set procedures to follow: let's try a few out and see if we like the field that results.

The somewhat grandiose rhetoric of the above paragraph could perhaps give the impression that this is the feel-good status quo option, an admission that all we need to do is to (continue) to think deeply without making any concrete changes. However, I think that this theory-building project is actually the \textit{most} radical. There are entire long-standing modes of visualization research that we would have to reconsider or even leave behind as part of this project, in the same way that a medieval alchemist would have little to contribute to a modern chemistry journal. A strong shared basis for visualization work would require a dramatic change in our self-conception and our policies, from the very first lecture of a visualization course to the instructions provided to reviewers to the papers or researchers acknowledged in test of time awards decades down the line.

Without shared projects, we risk irrelevance, fragility, and absurdity. Nor can our aspirations around plurality be used as a shield to put off the hard work of theory-building. In fact, a lack of theoretical structure risks the worst of both worlds in terms of basic and applied research: where the methods we use to justify our claims are illegible for the forms of knowledge we want to generate, and not credible for the people to whom we want to communicate this knowledge.

I note that the first step in such a theory-building epistemic project would be to assess its \textit{desirability}, let alone its \textit{feasbility} within the current milieu. Whether some abstract under-defined collective like ``the field'' is making progress through meta-analysis or theorizing might tug at your heartstrings less than more concrete questions about progress or flourishing for your users, your students, or yourself. The process of coming to know ourselves laid out in this section could, somewhat paradoxically, move us to a field where we simply do not care about scientific progress. For instance, while the VIS of the future could look more like a science, it could also end up resembling something like an art exhibition, where bold talents compare techniques or cohere around stylistic or aesthetic goals. Or the VIS of the future could be entirely subsumed into reporting on the new or changing needs and goals of our users and how existing techniques might help them, with ideas of innovating or systematizing visualization as a thing of the past. There are parts of those models and more that could be attractive. But I would like the field to (continue to) have some intentionality in its self-conception and direction.

\acknowledgments{
I wish to thank Miriah Meyer, Arvind Satyanarayan, and Vidya Setlur for their review and input on the survey design and participant list. I also wish to thank all three of them in addition to Steve Haroz for discussions on potential goals or format of this work, my participants for their thoughtful and generous engagement, and the anonymous reviewers for their feedback.}

\bibliographystyle{abbrv-doi}

\bibliography{template}
\end{document}